\documentclass[%
reprint,superscriptaddress,amsmath,amssymb,aps,
]{revtex4-2}

\usepackage[main=english,german,ngerman]{babel}

\usepackage{xcolor,colortbl}
\usepackage{braket}
\usepackage{dsfont}
\usepackage{hyphenat}
\usepackage{array}
\usepackage[export]{adjustbox}
\usepackage{amsmath,amssymb,amsfonts}
\usepackage[export]{adjustbox}
\usepackage{graphicx,wrapfig,lipsum}
\usepackage{graphicx}
\usepackage{dcolumn}
\usepackage{bm}
\usepackage{siunitx}

\usepackage{booktabs}
\usepackage{mathtools}
\usepackage{caption}
\usepackage{subcaption}
\usepackage[colorlinks=true,linkcolor=blue,urlcolor=blue,citecolor=blue]{hyperref}
\hypersetup{colorlinks}
\usepackage[capitalise]{cleveref}

\usepackage{physics}
\usepackage{bbold}
\usepackage{float}
\usepackage{multirow}
\usepackage[utf8]{inputenc}
\usepackage[toc,page]{appendix}

\usepackage{xspace}

\usepackage{suffix}
\usepackage{xstring}
\usepackage[nolist, nohyperlinks]{acronym}

\usepackage{comment}

\newcommand{\eg}{e.g.\xspace}
\newcommand{\ie}{i.e.\xspace}
\newcommand{\wrt}{with respect to\xspace}
\newcommand{\water}{\ensuremath{\text{H}_2 \text{O}}\xspace}

\newcommand{\MSE}[2]{\ensuremath{\text{MSE}(#1)_\text{#2}}\xspace}
\newcommand{\sutwo}{\ensuremath{\text{SU(2)}}\xspace}
\newcommand{\sothree}{\ensuremath{\text{SO(3)}}\xspace}
\newcommand{\stwo}{\ensuremath{\text{S}_2}\xspace}
\newcommand{\sn}{\ensuremath{\text{S}_n}\xspace}

\newcommand{\initstate}{\ensuremath{|\psi_0\rangle}}
\newcommand{\enclayer}[1]{\ensuremath{\Phi(#1)}}
\newcommand{\enclayeri}[2]{\ensuremath{\Phi^{(#1)}(#2)}}
\newcommand{\trainlayer}[1]{\ensuremath{\mathcal{U}_{#1}(\vec{\theta}_{#1})}}
\newcommand{\RH}[2]{\ensuremath{RH^{(#1)}(#2)}}
\newcommand{\encangle}{\ensuremath{\alpha_\text{enc}}}
\newcommand{\encanglea}[1]{\ensuremath{\alpha_{\text{enc},{#1}}}}
\newcommand{\SWAP}[1]{\ensuremath{\text{SWAP}^{(#1)}}}
\newcommand{\Hop}[1]{\ensuremath{H^{(#1)}}}

\newcommand{\sbanglei}[1]{\ensuremath{\varepsilon_{#1,\text{sb}}}}

\newcommand{\mc}[1]{\ensuremath{\mathcal{#1}}}

\begin{document}
\hypersetup{colorlinks=true,citecolor=magenta,linkcolor=magenta,filecolor=magenta,urlcolor=magenta,breaklinks=true}
\newcommand{\orcidicon}[1]{\href{https://orcid.org/#1}{\includegraphics[height=\fontcharht\font`\B]{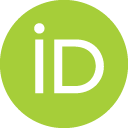}}}
\preprint{APS/123-QED}
\title{Symmetry-invariant quantum machine learning force fields}

\author{Isabel~Nha~Minh~Le\orcidicon{0000-0001-6707-044X}}
\email{isabel.le@rwth-aachen.de}
\affiliation{IBM Quantum, IBM Research Europe – Zurich, 8803 Rueschlikon, Switzerland}
\affiliation{
Institute for Quantum Information, RWTH Aachen University, 52074 Aachen, Germany
}

\author{Oriel~Kiss\orcidicon{0000-0001-7461-3342}}
\affiliation{European Organization for Nuclear Research (CERN), 1211 Geneva, Switzerland}
\affiliation{Department of Nuclear and Particle Physics, University of Geneva, 1211 Geneva, Switzerland}
\author{Julian~Schuhmacher\orcidicon{0000-0002-7011-6477}}
\affiliation{IBM Quantum, IBM Research Europe – Zurich, 8803 Rueschlikon, Switzerland}
\author{Ivano~Tavernelli\orcidicon{  0000-0001-5690-1981}}
\affiliation{IBM Quantum, IBM Research Europe – Zurich, 8803 Rueschlikon, Switzerland}
\author{Francesco~Tacchino\orcidicon{0000-0003-2008-5956}} 
\email{fta@zurich.ibm.com}
\affiliation{IBM Quantum, IBM Research Europe – Zurich, 8803 Rueschlikon, Switzerland}

\date{\today}

\begin{abstract}
Machine learning techniques are essential tools to compute efficient, yet accurate, force fields for atomistic simulations.
This approach has recently been extended to incorporate quantum computational methods, making use of variational quantum learning models to predict potential energy surfaces and atomic forces from \textit{ab initio} training data.
However, the trainability and scalability of such models are still limited, due to both theoretical and practical barriers.
Inspired by recent developments in geometric classical and quantum machine learning, here we design quantum neural networks that explicitly incorporate, as a data-inspired prior, an extensive set of physically relevant symmetries.
We find that our invariant quantum learning models outperform their more generic counterparts on individual molecules of growing complexity.
Furthermore, we study a water dimer as a minimal example of a system with multiple components, showcasing the versatility of our proposed approach and opening the way towards larger simulations.
Our results suggest that molecular force fields generation can significantly profit from leveraging the framework of geometric quantum machine learning, and that chemical systems represent, in fact, an interesting and rich  playground for the development and application of advanced quantum machine learning tools.
\end{abstract}

\maketitle

\begin{figure}[t]
    \centering
    \includegraphics[width=\columnwidth]{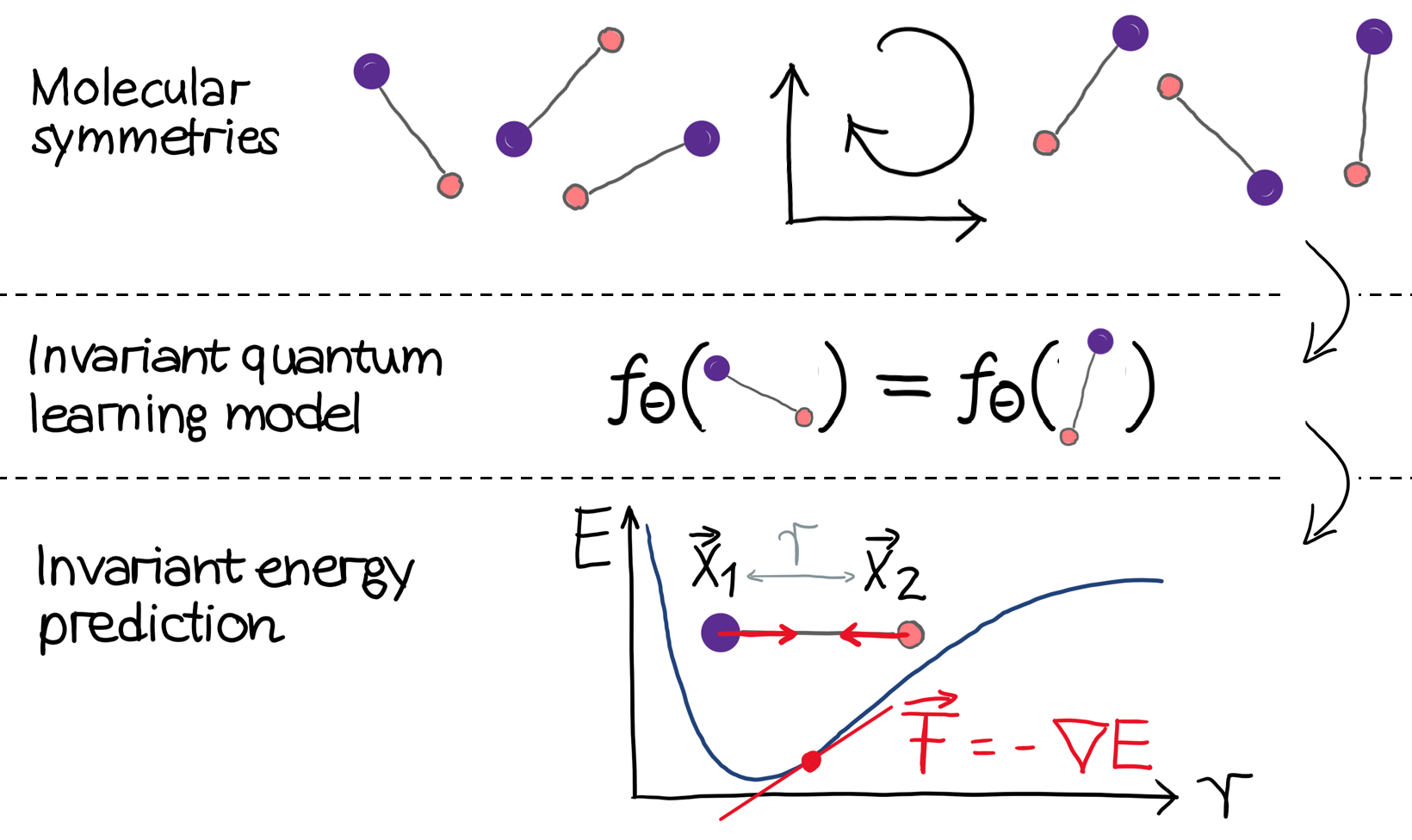}
    \caption{Overview of the work: we design invariant quantum learning models for a set of relevant molecular symmetries, obtaining -- upon input of simple Cartesian coordinates -- invariant predictions for potential energy surfaces and force fields to be employed in molecular dynamics simulations.
    }
    \label{fig:overview}
\end{figure}

\section{Introduction}
Atomistic simulations are essential computational tools for a wide range of research fields such as chemical physics, materials science, or biophysics~\cite{alder1959studies,mccammon1977dynamics,frenkel1997understanding}. 
Molecular dynamics -- one of the most prominent representatives of these computer simulation methods -- investigates properties of molecular systems by numerically integrating the mechanical equations of motion for each component in the system.
To this end, accurate knowledge of potential energy surfaces and atomic forces is required. In fact, the precision with which these quantities can be computed critically determines the reliability of the simulation.

While for medium-sized systems high-accuracy forces can be obtained with the so-called \textit{ab initio} methods, such as density functional theory~\cite{ballone1988equilibrium, marx2009ab}, their computational cost does not allow for the investigation of larger systems. 
In this case, empirically parameterized force fields, which are less computationally demanding but also less precise, are employed~\cite{monticelli2013force}.
Classical machine learning methods have been successfully exploited to learn the mapping between chemical configurations and the corresponding atomic forces by handling the task as a mathematical regression problem, obtaining a good balance between computational efficiency and prediction accuracy~\cite{behler2007generalized,cheng2020evidence,unke2021machine}. 
Recently, this philosophy has been extended to the realm of quantum machine learning (QML)~\cite{Kiss_2022} by employing \acp{VQLM} based on quantum neural networks~\cite{Mangini_2021,cerezo2021variational}. This approach gives access to a previously unexplored class of models that are, in principle, particularly well suited to capture genuine quantum mechanical properties. However, despite some promising results on small-scale examples, the applicability of such QML tools to problems of practical relevance is still limited. In fact, the scalability of \acp{VQLM} is often hindered by theoretical and practical challenges in trainability and generalization, which can be influenced by factors such as the choice of cost function, the presence of noise or, importantly, the expressivity of generic quantum circuit ansätze~\cite{mcclean2018barren,cerezo2021cost,holmes2022connecting,kubler2021inductive,NoiseBP,cerezo2022challenges,thanasilp2023subtleties,rudolph2023trainability,thanasilp2022exponential, unified_BP}. To specifically address some of these challenges, the use of meaningful inductive biases, often using carefully designed ansätze and losses, has been proposed and investigated~\cite{sauvage2022building,gruver2022deconstructing,PhysRevResearch.3.033090,Larocca2022diagnosingbarren,bowles2023contextuality,letcher2023tight,Melo2023pulseefficient,sannia2023engineered}. In the context of machine learning force fields, such strategy can be directly put into practice by considering relevant classes of molecular symmetries. 

In classical machine learning approaches~\cite{behler2007generalized}, and similarly, in the first QML studies~\cite{Kiss_2022}, native symmetries of the chemical systems under investigation are often accounted for by using invariant atomic descriptors as classical inputs. However, constructing unambiguous maps between molecular configurations and symmetry-invariant representations is, in general, a non-trivial task~\cite{pozdnyakov2020incompleteness, nigam2023completeness, sym_function_miksh,sym_function_zhang}. More recently, geometric deep learning~\cite{bronstein2017geometric, bogatskiy2022symmetry} brought a change in perspective, shifting the focus towards the design of models that natively respect the relevant sets of symmetries and yielding significant improvements in accuracy, robustness, and transferability~\cite{batzner2023advancing, musaelian2023learning}.

In this article, we leverage a quantum version of geometric machine learning to incorporate, in a QML-native way, molecular symmetries into \acp{VQLM}, employing different classes of equivariant quantum neural networks~\cite{exploiting_symmetry,nguyen2022theory,sauvage2022building,larocca2022group,skolik2023equivariant,schatzki2022theoretical,chang2023approximately}. With such physically motivated constructions, we report significant improvements in trainability and generalization capabilities in comparison to more generic quantum models, specifically in the cases of (i)~a single \ac{LiH} molecule, (ii)~a single water (\water) molecule, and (iii)~a dimer of \water molecules.
Our results demonstrate the effectiveness of symmetry-invariant \acp{VQLM}, suggesting a clear path towards a broader and more effective use of QML techniques for molecular force fields generation. Above all, our work confirms that this domain represents an ideal test bed for the application of advanced QML methods.

\section{\label{sec:methods}Model and methods}
\subsection{\label{sec:siVQLM} Symmetry-invariant quantum learning models}

A generic class of \acp{VQLM} has been applied in previous works~\cite{Kiss_2022} to the problem of learning molecular potential energy surfaces and force fields. 
These \acp{VQLM} map a set of classically preprocessed internal, hence rototranslationally invariant, coordinates $\mc{\Tilde{X}}$ to functional values 
\begin{equation}\label{eq:qnn-function}
    f_{\Theta}(\mc{\Tilde{X}}) = \langle \psi_0 | M(\mc{\Tilde{X}}, \Theta)^\dagger \mathcal{O} M(\mc{\Tilde{X}}, \Theta) | \psi_0\rangle
\end{equation}
parameterized by a set of weights $\Theta = \{\vec{\theta}_d\}_{d=0}^D$. 
The energy prediction is given by $E=f_\Theta(\mc{\Tilde{X}})$, while the force predictions with respect to $\Tilde{\mathcal{X}}$ are directly obtained from the \ac{VQLM} as $\vec{F}=-\nabla f_\Theta(\mc{\Tilde{X}})$, for instance via the parameter-shift-rule~\cite{schuld2019evaluating}.
The trainable weights are optimized to obtain a suitable map matching the model's prediction on some reference dataset by minimizing a predefined loss function using a classical optimization routine such as ADAM~\cite{adam}.
Note that both energy and forces need to be rescaled to lie in the output range of the \ac{VQLM}.
In~\cref{eq:qnn-function}, $\mathcal{O}$ is called an observable, $M_\Theta(\mc{\Tilde{X}})$ a quantum neural network, and $\initstate$ an initial state.
More specifically, the quantum neural network can generally be constructed as a re-uploading ansatz~\cite{gil2020input,effect_of_data_encoding, PerezSalinas2020datareuploading} consisting of interleaved data encoding layers $\enclayer{\mc{\Tilde{X}}}$ and parametrized, trainable layers $\trainlayer{d}$ with $\vec{\theta}_d \in \Theta$: 
\begin{equation}
    M_\Theta(\mc{\Tilde{X}}) = \left[ \prod_{d=D}^1 \enclayer{\mc{\Tilde{X}}} \trainlayer{d} \right] \enclayer{\mc{\Tilde{X}}} \, ,
\end{equation}
where $D$ denotes the depth of the \ac{VQLM}.
It can be shown that the quantum reuploading model expresses truncated Fourier sums \cite{gil2020input,effect_of_data_encoding}, where the accessible frequencies are only determined by the encoding layers. 
This property has been used to prove that, under reasonable hypotheses, \acp{VQLM} behave as universal functional approximators~\cite{effect_of_data_encoding,PhysRevA.104.012405}. 

In this work, we design \acp{siVQLM} \cite{exploiting_symmetry,nguyen2022theory} under certain symmetry groups relevant in the context of molecular dynamics. We will specifically leverage the framework and some of the results originally described in Ref.~\cite{exploiting_symmetry}, which we briefly summarize below.

A \ac{VQLM} is invariant under transformations from a group $G$ if it outputs the same label for all inputs $\mc{X}$ that are connected by a symmetry transformation at the data level. 
In the case of a reuploading model, this means that
\begin{equation}
    f_{\Theta}(V_g[\mc{X}]) = f_{\Theta}(\mc{X}) \quad \forall g\in G,
\end{equation}
where $V_g:G\rightarrow GL(\mathds{R}^3)$ is the representation of a certain element $g\in G$. The corresponding representation on the Hilbert space is denoted as $R_g:G\rightarrow GL(\mc{H})$.
Note that in this case, no further classical preprocessing of the input data is required, in contrast to the original \ac{VQLM} of Ref.~\cite{Kiss_2022}.

A $G$-invariant re-uploading model can be designed by choosing a $G$-invariant observable $R_g\mc{O}R_g^\dagger=\mc{O}$, a $G$-invariant initial state $R_g\initstate=\initstate$, and by additionally constructing the ansatz $M_\Theta(\mc{X})$ to be a $G$-equivariant quantum neural network, \ie, $M_\Theta(V_g[\mc{X}]) = R_g M_\Theta(\mc{X}) R_g^\dagger$. 
In other words: embedding a symmetry-transformed input vector $V_g[\mc{X}]$ results in the same parameterized quantum circuit as embedding the original input vector $\mc{X}$ and letting the symmetry transformation act at the quantum circuit level.
To this end, $M_\Theta(\mc{X})$ is built upon $G$-equivariant encoding and trainable layers defined by $\enclayer{V_g[\mc{X}]} = R_g \enclayer{\mc{X}} R_g^\dagger$ and $\left[\trainlayer{d}, R_g \right]=0$, respectively.
The required equivariant operations can be constructed using various tools such as the Twirling method, the Null space method, or the Choi operator method~\cite{exploiting_symmetry,nguyen2022theory}.
Since such \acp{siVQLM} do not depend on symmetry-invariant data inputs, a simple representation of the atomic configuration given by the Cartesian coordinates of each component in the system is sufficient, leading to minimal required effort in classical data preprocessing.

\subsection{\label{sec:relevant_symmetries}Relevant symmetries}
We consider the following three cases:
(i) A single molecule of two atoms of distinct species, \eg, \ac{LiH},
(ii) a single triatomic molecule of two atom types, \eg, \water,
and (iii) two molecules (dimer) of a triatomic molecule of two atom types, \eg, an \water dimer.
Molecular systems are invariant under the Euclidean symmetry group $E(3)$, \ie, global translations, rotations, and reflections.
Notice that the latter can be expressed as a combination of a translation and rotation in the cases of single molecules (i)-(ii), but it must be taken into account separately in the case of a dimer (iii).
Finally, systems composed of atoms of the same type possess an additional permutation symmetry. 
In this work, we respect translational symmetry by suitable choices of the reference Cartesian systems, as will be explained in the following subsections.
The remaining symmetries, \ie, rotation, and permutation of $n$ identical atoms or molecules are mathematically given by the groups \sutwo/\text{SO}(3) and \sn, respectively.
Additionally, a general reflection on the Euclidean space can be decomposed into a translation, a rotation, and a reflection along an arbitrary but fixed plane. Here, we choose this plane to be $(1, 1, 1)^\top$ leading to a data transformation of a Cartesian coordinate given by $\vec{x}\mapsto -\vec{x}$.

\subsection{\label{sec:equivariant_invariant_building_blocks}Equivariant and invariant building blocks}
In this section, we present some equivariant and invariant building blocks, which will be used to define \acp{siVQLM} in the remainder of the paper. 

\paragraph{\sutwo- and \stwo-invariant state.}\label{sec:singlet_state}
The singlet state of two qubits
\begin{align}\label{eq:singlet-state}
    |S\rangle &= 
    \frac{1}{\sqrt{2}} \left( |01\rangle - |10\rangle \right)
\end{align}
is invariant under rotations and qubit permutations. 
To see this, we first note that interchanging the two qubits leaves the quantum state unchanged, up to a global phase. 
Furthermore, a state $|S\rangle$ is rotationally invariant if it remains identical in any rotated orthogonal basis $\{u, u^\bot\}$, \ie, $|\psi_s\rangle = \frac{1}{\sqrt{2}}\left( |u u^\bot\rangle - |u^\bot u\rangle \right)$.
By writing $|u\rangle = \alpha |0\rangle + \beta |1\rangle$ with $\alpha,\beta\in\mathds{C}$,  $|\alpha|^2+|\beta|^2=1$ and $|u^\bot\rangle = -\beta^\ast |0\rangle + \alpha^\ast |1\rangle$ such that $\langle u|u^\bot\rangle = \langle u^\bot|u\rangle =0$ and rearranging the terms, it can be seen that indeed $|uu^\bot\rangle - |u^\bot u\rangle = |01\rangle - |10\rangle$.
For any even number of $2m$ of qubits, a rotationally invariant state $|\psi\rangle$ can be obtained with the tensor product of $m$ copies of the singlet state, \ie, $|\psi\rangle = \bigotimes_{i=1}^m|S\rangle$.
In the following, $|S_{ij}\rangle$ denotes a singlet state on a pair of qubits $(i,j)$.

\paragraph{\sutwo-invariant and -equivariant operators.}
Let $\vec{\sigma}^{(i)} = \left(X^{(i)}, Y^{(i)}, Z^{(i)}\right)^\top$ denote the three-dimensional vector of Pauli operators acting on qubit $i$.
To find a rotationally invariant operator we start from the rotationally invariant Heisenberg Hamiltonian $H_\text{Heis}(J) = -J \sum_{(i,j)} \vec{S}^{(i)} \vec{S}^{(j)}$, where $(i,j)$ denotes pairs of qubits, $J\in\mathds{R}$, and $\vec{S}=\frac{\hbar}{2}\vec{\sigma}$, is invariant under rotation.
Based on this, we propose a two-qubit operator acting on a pair of qubits $(i,j)$ that is invariant under rotations of the overall system, and invariant under swapping of qubits $i$ and $j$, as
\begin{align}\label{eq:heisenberg_interaction}
H^{(i,j)}(J) &= J \, \vec{\sigma}^{(i)} \vec{\sigma}^{(j)} \, ,\quad J\in\mathds{R}.
\end{align}
An overall system rotation about axis $l\in\{X,Y,Z\}$ is generated by the corresponding component of the total Pauli operator of $N\geq 2$ qubits, $\sigma_l^\text{tot} = \sum_{i=1}^{N} \sigma_l^{(i)}$, \ie,
\begin{align}
    R_l(\varphi) &= \exp\left(-i \frac{\varphi}{2} \sigma_l^\text{tot}\right) \, ,\quad \varphi\in\mathds{R}.
\end{align}
Since we have that $\left[\sigma_l^\text{tot}, \vec{\sigma}^{(i)}\vec{\sigma}^{(j)}\right]=0$ for all $1\leq i,j \leq N$, the operator given in \cref{eq:heisenberg_interaction} is invariant under \sutwo.
We call this operator the Heisenberg interaction between qubits $(i,j)$.
Finally, an \sutwo-equivariant trainable block on two qubits $(i,j)$ is obtained by
\begin{align}\label{eq:RH_equivariant}
    RH^{(i,j)}(J) = \exp \left( -i H^{(i,j)}(J) \right),
\end{align}
where $J\in\mathds{R}$ is a trainable parameter.

\paragraph{\sutwo-equivariant encoding block.}
We make use of the \sutwo-equivariant embedding of a single data point $\vec{x}\in\mathds{R}^3$ on qubit $i$ introduced in Ref.~\cite{exploiting_symmetry} given by $\Phi^{(i)}(\vec{x}) = \exp\left( -i \cdot \vec{x} \vec{\sigma}^{(i)} \right)$, and further extend it with a scaling factor $\encangle\in\mathds{R}$ to gain additional expressivity:
\begin{align}\label{eq:EQNN-single-encoding}
    \enclayeri{i}{\vec{x};\alpha_\text{enc}} = \exp\left( -i\alpha_\text{enc} \cdot \vec{x} \vec{\sigma}^{(i)} \right).
\end{align}
In the following, we sometimes omit $\encangle$ for better readability, \ie, $\Phi^{(i)}(\vec{x};\alpha_\text{enc}) = \Phi^{(i)}(\vec{x})$.
Since the only difference between \cref{eq:EQNN-single-encoding} and the equivariant embedding given in Ref.~\cite{exploiting_symmetry} is a scaling factor, the proposed data embedding is still \sutwo-equivariant.
To see this, consider single-qubit operators and a rotation $r\in\text{SO}(3)$ that we decompose into three canonical rotations $r(\psi, \theta, \phi)=r_z(\psi)r_x(\theta)r_z(\phi)$. Here, $r_i(\varphi)$ denotes a rotation of an angle $\varphi\in\mathds{R}$ with respect to axis $i$. Since $\left( r_i(\varphi)\vec{x}\right) \vec{\sigma} = R_i(-\varphi) \vec{x} \vec{\sigma}R_i(\varphi)$ \cite{exploiting_symmetry}, where $R_i(\varphi)$ is the single-qubit rotation gate corresponding to $r_i$, we have that $\Phi(r(\psi,\theta,\phi)\vec{x}, \alpha_\text{enc}) = \mc{R}_g \Phi(\vec{x}, \alpha_\text{enc}) \mc{R}_g^\dagger$, where $\mc{R}_g = R_Z(\psi)R_X(\theta)R_Z(\phi)$ represents a general rotation operation on the qubit space.

\paragraph{Reflection-equivariant encoding block.}
As shown in Ref.~\cite{exploiting_symmetry}, the embedding given in \cref{eq:EQNN-single-encoding} can be further extended to be equivariant under reflection with respect to $(1, 1, 1)^\top$ by making use of an additional qubit $j\neq i$:
\begin{equation}\label{eq:reflection-equivariant-embedding}
    \enclayeri{i,j}{\vec{x};\alpha_\text{enc}}= \exp\left( -i\alpha_\text{enc} \cdot \vec{x} \vec{\sigma}^{(i)} X^{(j)} \right).
\end{equation}
To see this, we note that $\Phi^{(i,j)}(-\vec{x}) = Z^{(j)} \Phi^{(i,j)}(\vec{x}) Z^{(j)}$.

\subsection{Diatomic molecule: a quantum learning model invariant under rotations}
As a first step, we make use of the building blocks described above to construct an \sutwo-invariant \ac{VQLM} defined on $N=4$ qubits for a diatomic system, to be applied to the \ac{LiH} molecule.
Leveraging the flexibility of quantum reuploading models, we will encode the position of each atom twice for higher expressivity.
The model inputs the Cartesian positions $\mc{X}=(\vec{x}_1, \vec{x}_2)$ of both atoms and outputs \sutwo-invariant energy predictions, as well as the atomic forces with respect to Cartesian coordinates.
The translational invariance is guaranteed by centering the positions of the two atoms around the origin, \ie, $\vec{x}_1 = -\vec{x}_2$, and hence using translational invariant inputs.
We extend the embedding given by \cref{eq:EQNN-single-encoding} to encode two data points $(\vec{x}_1, \vec{x}_2)$ into four qubits by means of
\begin{align}\label{eq:LiH-EQNN-encoding}
    \Phi(\vec{x}_1, \vec{x}_2) = \Phi^{(1)}(\vec{x}_1) \Phi^{(2)}(\vec{x}_2) \Phi^{(3)}(\vec{x}_1) \Phi^{(4)}(\vec{x}_2),
\end{align}
where for better readability we used $\enclayeri{i}{\vec{x}}=\enclayeri{i}{\vec{x};\encangle}$.
As discussed in the previous section, representations of the rotational group on $\mathds{R}^3$ can be constructed by $V_g = V_{\psi,\theta,\phi} = r_X(\psi)r_Z(\theta)r_X(\phi)$, where $r_l(\varphi)\in \sothree$ are rotation matrices.
On the Hilbert of a single qubit $i$, an equivalent representation is given by the corresponding three single-qubit rotations $R_g^{(i)} = R_{\psi,\theta,\phi}^{(i)} = R_X^{(i)}(\psi)R_Z^{(i)}(\theta)R_X^{(i)}(\phi)$.
This can be extended to four qubits using the tensor representation~\cite{nguyen2022theory} $R_{\psi,\theta,\phi}^{\otimes 4}=\prod_{i=1}^4 R_{\psi,\theta,\phi}^{(i)}$.
In this way, it can easily be seen that the embedding in \cref{eq:LiH-EQNN-encoding} is still equivariant under \sutwo.

The \sutwo-equivariant parameterized operator given by \cref{eq:RH_equivariant} can be extended to four qubits by
\begin{widetext}
\begin{equation}\label{eq:RH_four}
    \mc{U}_d(\vec{j}_d) = \prod_{b=1}^B \left[\RH{1,2}{j_1^{db}} \RH{3,4}{j_2^{db}} \RH{2,3}{j_3^{db}}\right] \quad\quad 0\leq d \leq D,
\end{equation}
\end{widetext}
where $\vec{j}_d\in\mathds{R}^{B\times 3}$ is the vector of weights in the trainable layer $d$, and $B$ is another hyperparameter that can be manually tuned to increase the expressivity of the circuit.

The rotation\hyp invariant two\hyp qubit singlet state given in \cref{eq:singlet-state} is extended to a four-qubit initial state as:
\begin{align}\label{eq:invariant-state-4}
    |\psi_0\rangle &= |S_{12}\rangle \otimes |S_{34}\rangle.
\end{align}

\begin{figure*}[t]
    \centering
    \includegraphics[width=\textwidth]{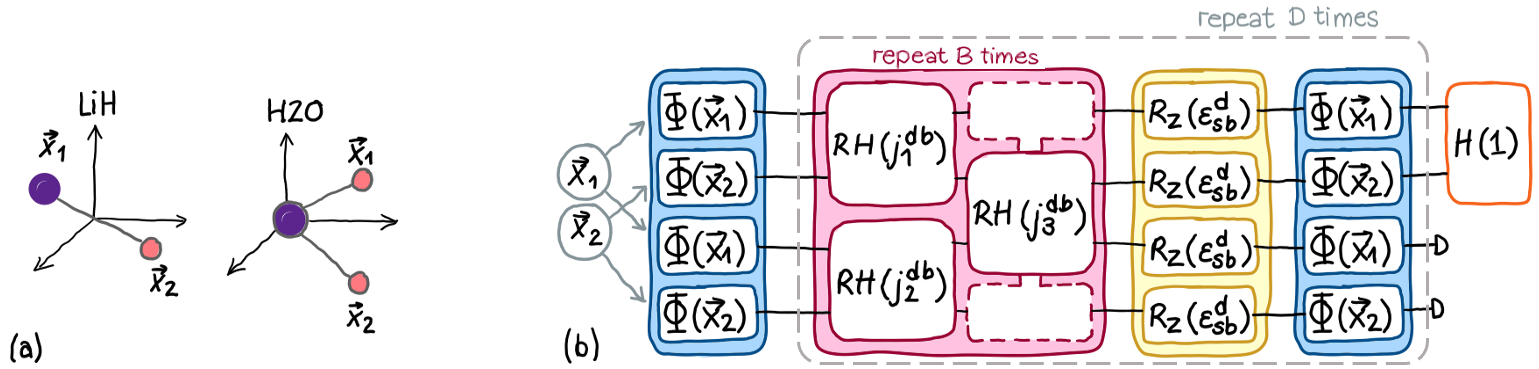}
    \caption{Symmetry-invariant models for diatomic (\ac{LiH}) and triatomic (\water) molecules: (a)~ Translationally invariant inputs are obtained by fixing the origin of the Cartesian coordinate system. For \ac{LiH}, we choose $\vec{x}_1 = -\vec{x}_2$, while for \ac{H2O} $\vec{x}_O =\vec{0}$ and, in general, $\vec{x}_1 \neq -\vec{x}_2$, where $\vec{x}_1, \vec{x}_2$ are the coordinates of the hydrogen atoms. (b)~The \ac{siVQLM} for a single molecule of \ac{LiH} and \ac{H2O}. The equivariant embedding layers (blue) encode $\mc{X}=(\vec{x}_1, \vec{x}_2)$ via \cref{eq:LiH-EQNN-encoding}. Importantly, note that this embedding depends on a trainable parameter $\alpha_\text{enc}\in\mathds{R}$. The equivariant trainable layers (red) are given by \cref{eq:RH_four} and \cref{eq:h2O_trainable}. In the case of \ac{H2O}, the trainable layer is extended with the red dashed operator and an optional symmetry-breaking layer (yellow). Finally, the invariant observable $\mc{O}$ (orange) is measured.}
    \label{fig:model}
\end{figure*}

Based on the \sutwo-invariant operator given in \cref{eq:heisenberg_interaction}, an invariant observable can be chosen as
\begin{equation}\label{eq:invariant_observable_choice}
    \mc{O} = \vec{\sigma}^{(1)}\vec{\sigma}^{(2)}.
\end{equation}
Notice that the output of this \ac{siVQLM} lies in the range of $[-1, 3]$: therefore, the energy and force labels need to be rescaled accordingly, for instance using a MinMaxScaler. 
The overall \sutwo-invariant \ac{VQLM} for a diatomic molecule is visualized in \cref{fig:model}.

\subsection{Triatomic molecule of two atom types: a quantum learning model invariant under rotations and permutation}\label{sec:VQLM-H2O}
For a triatomic molecule with two atomic species, the previous \ac{siVQLM} can be modified to additionally consider invariance under the permutation of identical atoms.
A specific example of such a molecular system is given by a single \water molecule.
The unique atom is set in the origin of the Cartesian coordinate system. This reduces the dimension of the input data to two Cartesian coordinates, $\mc{X}=(\vec{x}_1,\vec{x}_2)$, representing the position of the two hydrogen atoms. Fixing the origin to the unique atom of oxygen also ensures translational invariance.
In contrast to the diatomic case, note that in general $\vec{x}_1 \neq -\vec{x}_2$. 
The permutational invariance of identical atoms is already fulfilled by the embedding defined in~\cref{eq:LiH-EQNN-encoding}.
Interchanging identical atoms is represented on the Hilbert space of four qubits by swapping the corresponding qubits, $R=\SWAP{1,2} \SWAP{3,4}=R^\dagger$, from which it can easily be seen that the embedding is indeed \stwo-equivariant.

The full \stwo-equivariant trainable layer can then be written as:
\begin{widetext}
\begin{align}\label{eq:h2O_trainable}
    \mc{U}_d(\vec{j}_d) = \prod_{b=1}^B \left[\RH{1,2}{j_1^{db}} \RH{3,4}{j_2^{db}} \RH{2,3}{j_3^{db}} \RH{1,4}{j_3^{db}}\right] \quad\quad 0\leq d \leq D,
\end{align}
\end{widetext}
where $\vec{j}_d\in \mathds{R}^{B\times 3}$ are trainable weights.
By expressing the SWAP-operator in terms of Pauli operators, \ie, $\SWAP{i,j} =\frac{1}{2} \sum_{P}P^{(i)} P^{(j)}$ for $P\in\{\mathds{1}, X, Y, Z\}$,
it can be shown that $\SWAP{i,j}$ commutes with each Heisenberg interaction given in \cref{eq:heisenberg_interaction} acting on either the same pair of qubits or any disjoint one, i.e., $[\text{SWAP}(i,j), H^{(k,l)}]=0$ if and only if $(i,j)=(k,l)$ or the two pairs do not have any qubit in common. Moreover, it holds that $[\SWAP{i,j}, \Hop{i,l}] + [\SWAP{i,j}, \Hop{l,j}] = 0$ for $i\neq l\neq j$, where in both commutators there is a shared qubit ($i$ in the former and $j$ in the latter) between $\SWAP{i,j}$ and the Heisenberg interaction term.
As a result, the introduced \stwo-representation commutes with all generators of each component of the product in \cref{eq:h2O_trainable}, namely $\Hop{1,2}$, $\Hop{3,4}$, and $\Hop{2,3}\Hop{1,4}$, implying \stwo-equivariance of \cref{eq:h2O_trainable}.
Furthermore, this analysis also shows that \cref{eq:invariant_observable_choice} is not only \sutwo-invariant but also \stwo-invariant and hence represents a suitable choice for an observable in this case.
Since the singlet state is an eigenstate of the SWAP-operator with eigenvalue $-1$, $\initstate$ as given in \cref{eq:invariant-state-4} is also \stwo-invariant: $\SWAP{1,2}\SWAP{3,4}\initstate = \initstate$.

It is now worth mentioning that while using symmetry-respecting ansätze restricts the expressivity in a meaningful way,
this might sometimes lead to unfavorable loss landscapes~\cite{exploiting_symmetry}. 
In such cases, the trainability can potentially be improved by introducing a tunable unitary $\mc{U}_{d,\text{sb}}$, that controls the amount of symmetry-breaking in the mode~\cite{exploiting_symmetry,park2021efficient}. To test this idea, we extend each trainable layer by a symmetry-breaking unitary, 
\begin{equation}\label{eq:symmetry-breaking}
    \mc{U}_{d,\text{sb}}(\varepsilon_{d,\text{sb}})=\prod_{i=1}^{N} R_Z^{(i)}(\sbanglei{d}) \quad\quad 0\leq d\leq D,
\end{equation}
parameterized by a trainable $\sbanglei{d}$ identical for all qubits in the considered layer $d$. 
Symmetry-breaking can be easily excluded by fixing $\sbanglei{d}=0$ for all $0\leq d\leq D$.

As in the previous case, the output of this \ac{siVQLM} lies in the range of $[-1, 3]$ such that the energy and force labels need to be rescaled accordingly.
The overall architecture is visualized in \cref{fig:model}.
\subsection{Triatomic dimer of two atom types: a quantum learning model invariant under rotations, permutations, and reflections}

\begin{figure*}
    \centering
    \includegraphics[width=\textwidth]{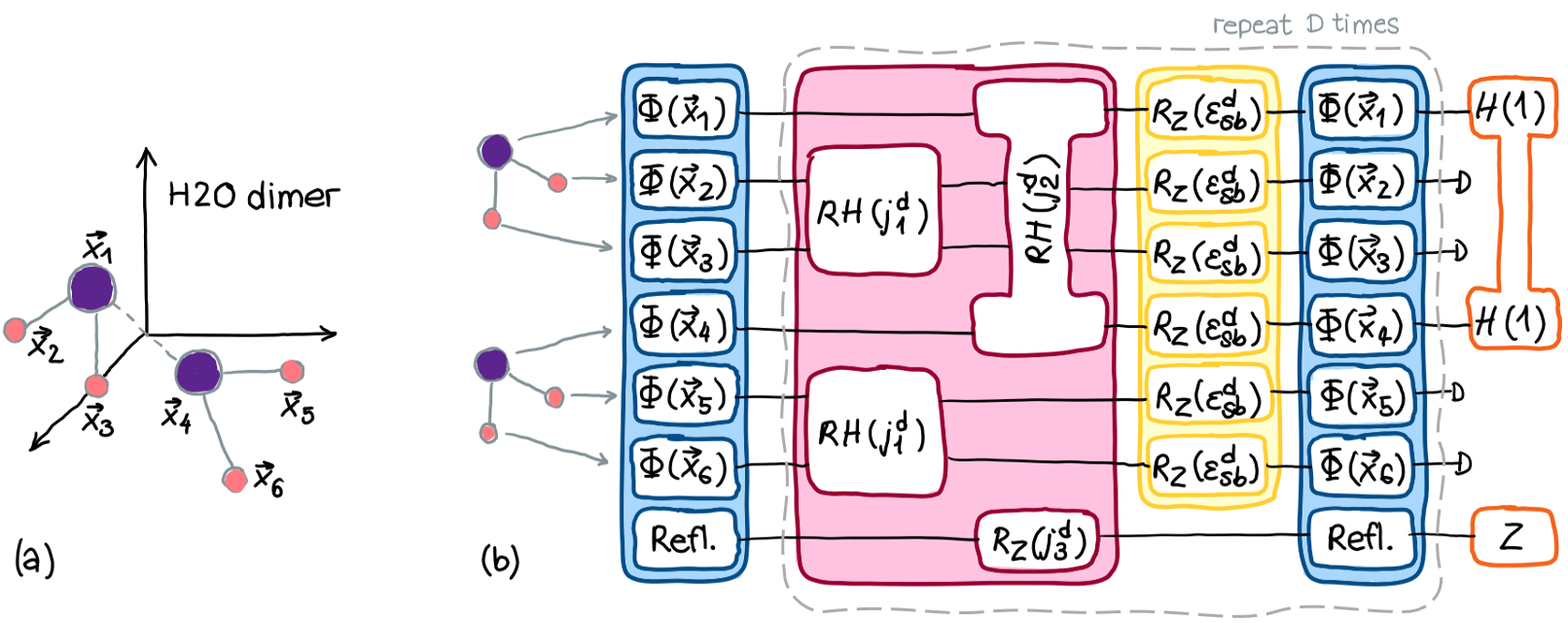}
    \caption{Symmetry-invariant model for the example of two \water molecules. (a)~Translationally invariant inputs are obtained by fixing the origin of the Cartesian coordinate system. The oxygen atoms lie on opposite sites of the origin, i.e., $\vec{x}_O^1 = -\vec{x}_O^2$. (b)~The \ac{siVQLM} for \water dimer. The equivariant embedding layers (blue) encode $\mc{X}=(\vec{x}_i)_{i=1}^6=(\vec{x}_O^1, \vec{x}_H^1, \vec{x}_H^1, \vec{x}_O^2, \vec{x}_H^2, \vec{x}_H^2)$ via \cref{eq:H2O_dimer_encoding}, where the embedding depends on a atom-type $a$ specific trainable parameter $\encanglea{a}\in\mathds{R}$. The equivariant trainable layers (red) are given by \cref{eq:h2O_dimer_trainable} and extended by a symmetry-breaking layer (yellow). In the end, the invariant observable $\mc{O}$ (orange) is measured.}
    \label{fig:H2O_dimer_siVQLM}
\end{figure*}

As a final step, the complexity is increased by considering a dimer of triatomic molecules with two atom types.
A specific example of such a molecular system is a \water dimer.
In this case, reflections with respect to a plane need to be considered explicitly together with rotations and translations. 
Furthermore, the model has to respect the interchanging of both molecules.
Notice that the design of an \ac{siVQLM} for a system of two molecules is an important step towards the application of QML methods to larger molecular systems, where intermolecular interactions must be captured. 
In the following, we present an \ac{siVQLM} based on $N=7$ qubits respecting all the mentioned symmetries.
As argued in \cref{sec:relevant_symmetries}, it is sufficient to consider reflections with respect to the plane $(1, 1, 1)^\top$, \ie, $V[\vec{x}]=-\vec{x}$.

Translational invariance is once again incorporated by fixing the origin in the data space.
To this end, the two unique atoms of the molecule are centered around the origin, \eg, for \water: $\vec{x}_{O}^1= -\vec{x}_{O}^2$.
The input vector is then given by the set of Cartesian coordinates for each atom in the following order (in the case of \water):
$\mc{X}=(\vec{x}_i)_{i=1}^{N-1}=(\vec{x}_{O}^1, \vec{x}_{H1}^1, \vec{x}_{H2}^1, \vec{x}_{O}^2, \vec{x}_{H1}^2, \vec{x}_{H2}^2)$, where $\vec{x}_{ai}^j$ denotes the position of the $i$th atom of atom type $a$ in molecule $j$.

The embedding given by \cref{eq:reflection-equivariant-embedding} is extended to encode the six atoms as:
\begin{equation}\label{eq:H2O_dimer_encoding}
    \enclayer{\mc{X}} = \prod_{i=1}^{N-1}\exp\left( -i\encanglea{a} \vec{x}_{i} \vec{\sigma}^{(i)} X^{(N)}\right),
\end{equation}
where we make use of distinct trainable scaling factors $\encanglea{a}$ for each atom type $a$.
Using this embedding scheme, the reflection on the Hilbert space is given by its representation $\mc{R}=Z^{(N)}=\mc{R}^\dagger$, while a rotation is still represented by the corresponding single-qubit rotations.
The interchanging of the two identical atoms in the first (second) molecule is represented by \SWAP{2,3} (\SWAP{5,6}), and the permutation of both molecules has the representation $\SWAP{1,4}\SWAP{2,5}\SWAP{3,6}$.

We propose a trainable layer that is equivariant under rotation, relevant permutation, and reflection as:
\begin{widetext}
\begin{align}\label{eq:h2O_dimer_trainable}
    \mc{U}_d(\vec{j}_d) = \RH{2,3}{j_1^{d}} \RH{5,6}{j_1^{d}} \RH{1,4}{j_2^{d}} R_Z^{(7)}(j_3^{d}) \quad\quad 0\leq d \leq D,
\end{align}
\end{widetext}
where $\vec{j}_d\in\mathds{R}^3$, $RH^{(i,j)}$ is the Heisenberg interaction acting on a pair of qubits $(i,j)$ as defined in \cref{eq:heisenberg_interaction}, and $R_Z^{(i)}(\phi)$ is a $Z$-rotation of qubit $i$ about an angle $\phi\in\mathds{R}$.
Note that here we do not repeat the trainable block $B$ times as in the case of a single diatomic or triatomic molecule. This is because all trainable operations in \cref{eq:h2O_dimer_trainable} commute with each other, since they act on different (pairs of) qubits, hence any repetition would result in the same effective trainable layer with slightly different parametrization.
With an argument similar to the one presented in \cref{sec:VQLM-H2O}, the introduced representations of the relevant permutations commute with all generators of each of the product components in \cref{eq:h2O_dimer_trainable}, namely $H^{(2,3)} H^{(5,6)}$, and $H^{(1,4)}$.
Hence, the trainable layer proposed in \cref{eq:h2O_dimer_trainable} is equivariant under the relevant permutations.

An invariant observable is then given by,
\begin{equation}
    \mc{O}=H^{(1,4)} Z^{(7)}.
\end{equation}

An invariant initial state can be obtained by tensoring singlet states of suitable pairs of qubits:
\begin{equation}
    |\psi_0\rangle = |S_{23}\rangle\otimes |S_{56}\rangle \otimes |S_{41}\rangle \otimes |0\rangle.
\end{equation}

Similarly to the \water single molecule case, symmetry-breaking is introduced in each trainable layer by \cref{eq:symmetry-breaking}.
The output of this \ac{siVQLM} lies in the range of $[-4, 2]$, meaning that the energy and force labels need to be rescaled accordingly.
A sketch of the overall \ac{siVQLM} architecture is shown in \cref{fig:H2O_dimer_siVQLM}.

\section{\label{sec:results}Results}
\subsection{Diatomic molecule: lithium hydride}
\paragraph{Training on exact energy labels.}
\begin{table*}
\caption{\label{tab:LiH_results}
Numerical results for \ac{LiH}. For the original \ac{VQLM} a distinction is made between the full test set (including outliers) and the relevant test set (excluding outliers), while for the \ac{siVQLM} only the full test set is taken into account.}
\begin{ruledtabular}
\begin{tabular}{lccc}
\multicolumn{1}{c}{\textit{Lithium hydride}} & \ac{VQLM} (full) & \ac{VQLM} (relevant)& \ac{siVQLM} \\
\hline
$\MSE{E}{test}$ (eV$^2$) & $2.48 \cdot 10^{-2}$ & $1.20\cdot 10^{-6}$ & $1.89\cdot 10^{-6}$ \\
$\MSE{F}{test}$ (eV$^2$\AA$^{-2}$) & $24.91$ & $1.23 \cdot 10^{-3}$ & $3.14 \cdot 10^{-4}$ \\
\end{tabular}
\end{ruledtabular}
\end{table*}

\begin{figure}
    \centering
    \includegraphics[width=\columnwidth]{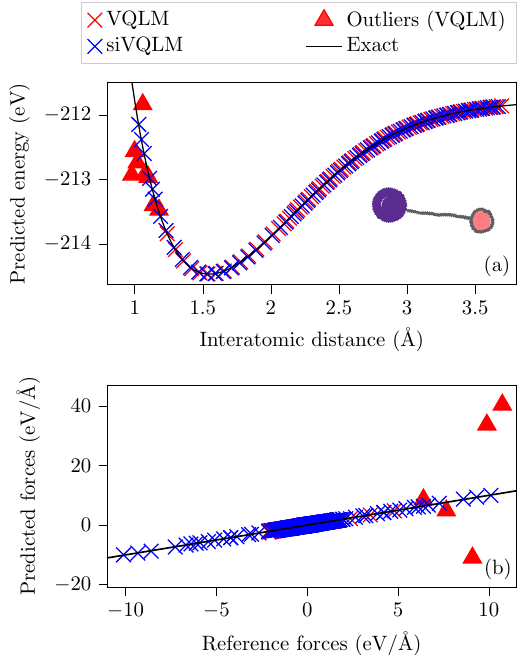}
    \caption{Energy prediction (a) and force prediction (b) for \ac{LiH}.
    The predictions by the generic \ac{VQLM} (red crosses) and the \ac{siVQLM} (blue crosses) are compared to the exact energy/forces (black solid line).
    The force prediction in (b) by the generic \ac{VQLM} is given with respect to internal coordinates, while the \ac{siVQLM} directly predicts atomic forces with respect to Cartesian coordinates.}
    \label{fig:LiH-results}
\end{figure}

We choose \ac{LiH} as a paradigmatic  diatomic molecule and compare the more generic \ac{VQLM} presented in Ref.~\cite{Kiss_2022} with our proposed \ac{siVQLM}.
The \ac{siVQLM} is built as described in \cref{sec:methods} with $B=1$ and $D=22$ resulting in $N_\text{params}=67$ trainable parameters, whereas the generic \ac{VQLM} is built using $D=13$ resulting in $N_\text{params}=94$ trainable weights.
Both models are initialized following Ref.~\cite{Grant2019initialization} to mitigate the potential occurrence of barren plateaus at the beginning of the training, and trained with a maximum of $N_\text{maxiter}=3000$ steps of the ADAM optimizer~\cite{adam}, where we choose the loss function to be the mean-squared-error of the energy labels.
The classical reference energies are constructed by numerically diagonalizing the second quantized Hamiltonian expressed in the STO3G basis set for bond lengths $r\in[0.9, 4.5]$ \AA, while the reference forces are computed via finite differences over the exact potential energy surface.
The \ac{siVQLM} is trained on $|\mc{A}_\text{train}|=53$ training points and tested on $|\mc{A}_\text{test}|=81$ points, while the original \ac{VQLM} is trained on a periodized dataset with $|\mc{A}_\text{train}|=113$ training points and tested on $|\mc{A}_\text{test}|=187$ following the procedure described in Ref.~\cite{Kiss_2022}.
The results for the energy and force prediction are visualized in \cref{fig:LiH-results} and given in \cref{tab:LiH_results}.

We can immediately observe that the \ac{siVQLM} makes precise predictions for previously unseen configurations over the entire range covered by the training set, while the original \ac{VQLM} fails in the higher energy range for short interatomic distances.
Since typically only the lower energy range is considered within molecular dynamics, we exclude the outliers depicted in \cref{fig:LiH-results} in the test set of the original \ac{VQLM} for better comparison with the \ac{siVQLM}.
Despite the exclusion of high energy points, the original model is outperformed by the \ac{siVQLM}. 
While the energy prediction of both \acp{VQLM} reach a comparable accuracy, the force prediction of the \ac{siVQLM} exhibits an improvement in accuracy of about one order of magnitude.
This indicates stronger generalization capacities when including underlying symmetries as an inductive bias of the \ac{VQLM}, which also leads to smoother energy predictions and, consequently, to significantly better forces.
Furthermore, the \ac{siVQLM} directly predicts the atomic forces \wrt Cartesian coordinates.

\begin{figure}[H]
    \includegraphics[width=\columnwidth]{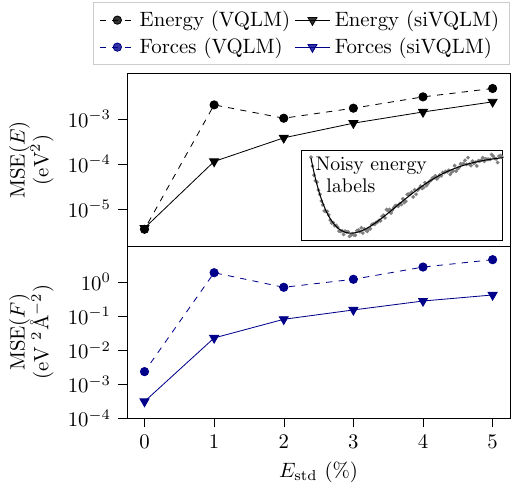}
    \caption{Training on noisy energy labels for \ac{LiH}. The energy (black) and force (blue) prediction of the \ac{VQLM} and \ac{siVQLM} are compared for increasing amounts of noise $E_\text{std}$. In the case of the original \ac{VQLM}, the loss is taken for the relevant test data set as before in the exact case. The inset shows the noisy energy labels for $E_\text{std}=0.05$.}
    \label{fig:LiH_noisy}
\end{figure}

\paragraph{Training on noisy energy labels.}
While the energy and force labels can, in practice, be obtained with high accuracy when utilizing 
%precise 
noiseless
first-principle techniques such as density functional theory, employing different methods to generate training labels, including near-term quantum computing techniques, is generally affected by statistical and systematic errors \cite{schuhmacher2022extending}. 
Studying their impact is therefore useful in view of building, for instance, a fully quantum-powered force field generation and learning workflow.
We investigate this possibility for both \acp{VQLM} discussed above, training them on identical data sets containing noisy energy labels with increasing amount of noise $E_\text{std}$.
The noisy labels are constructed from noiseless ones by adding random contributions drawn from a Gaussian distribution centered around $E_0 = 0$,
\begin{equation}
    \mc{N}_{E_\text{std}}(E) = \frac{1}{\sqrt{2\pi E_\text{std}^2}} \exp\left( - \frac{E^2}{2E_\text{std}^2} \right).
\end{equation}
An example for $E_\text{std}=0.05$ is visualized in the inset of \cref{fig:LiH_noisy}.
As before, the same relevant test batches are considered to quantify the prediction accuracy of the original \ac{VQLM}, while the full test data set is considered for the \ac{siVQLM}.
The results are visualized in \cref{fig:LiH_noisy}.

Not surprisingly, the prediction accuracy decreases with increasing noise for both models.
However, while in the noise-free case, the energy prediction of the \ac{VQLM} (only on the relevant test data set) and the \ac{siVQLM} were essentially comparable, the \ac{siVQLM} becomes superior when considering noisy training labels, especially for small amounts of noise. 
Concerning the atomic forces, the \ac{VQLM} fails to predict meaningful forces even when only the relevant test set (low-energy region) is taken into account.
The \ac{siVQLM} instead is superior by at least one order of magnitude, confirming the observations made in the noise-free case and showcasing a significantly improved robustness.

\begin{table*}
\caption{\label{tab:H2O_results}
Numerical results for \water of the original \ac{VQLM}, a non-symmetry-breaking \ac{siVQLM} (no SB), and a symmetry-breaking \ac{siVQLM} (SB).
}
\begin{ruledtabular}
\begin{tabular}{lccc}
\multicolumn{1}{c}{\textit{Water}} & \ac{VQLM} & \ac{siVQLM} (no SB) & \ac{siVQLM} (SB)\\
\hline
$\MSE{E}{train}$ (eV$^2$) & $1.69\cdot 10^{-6}$ & $2.19\cdot 10^{-6}$ & $4.59\cdot 10^{-7}$ \\
$\MSE{F}{train}$ (eV$^2$\AA$^{-2}$) & $3.29\cdot 10^{-3}$ & $1.36\cdot 10^{-3}$ & $2.10\cdot 10^{-4}$ \\
\hline
$\MSE{E}{test}$ (eV$^2$) & $3.84 \cdot 10^{-6}$ & $3.40\cdot 10^{-6}$ & $4.54 \cdot 10^{-7}$ \\
$\MSE{F}{test}$ (eV$^2$\AA$^{-2}$) & $5.55 \cdot 10^{-3}$ & $1.53\cdot 10^{-3}$ & $2.39 \cdot 10^{-4}$ \\
\end{tabular}
\end{ruledtabular}
\end{table*}
\subsection{Triatomic molecule composed of two atom types: water}

For the triatomic case with two atom types, we choose \water as a paradigmatic molecule and compare the original \ac{VQLM}~\cite{Kiss_2022} with the \ac{siVQLM}.
To test the idea of actively breaking some symmetries to obtain a smoother loss landscape, we distinguish between a non-symmetry-breaking \ac{siVQLM} and a symmetry-breaking model.
Practically, a non-symmetry-breaking \ac{siVQLM} is obtained by fixing all symmetry-breaking angles in \cref{eq:symmetry-breaking} to zero and excluding them from the set of trainable weights.
All models are built using $D=11$ encoding and trainable layers, where within the \ac{siVQLM} we choose $B=2$ and additionally $\varepsilon_\text{sb,initial}=0.1$ for all layers in the case of symmetry-breaking.
This leads to a \ac{VQLM} with $N_\text{params}=80$, a symmetry-breaking \ac{siVQLM} with $N_\text{params}=78$, and a non-symmetry-breaking \ac{siVQLM} with $N_\text{params}=67$ trainable weights.
The reference energy and force labels, which are computed using density functional theory, are retrieved from Ref.~\cite{morawietz2019hdnnp}.
The models are then each trained on $|\mc{A}_\text{train}|=473$ points and tested on $|\mc{A}_\text{test}|=474$ points with the same loss function and optimizer as for the diatomic example.
The numerical results on the energy and force prediction are listed in \cref{tab:H2O_results}.

\begin{figure}
    \centering
    \includegraphics[width=\columnwidth]{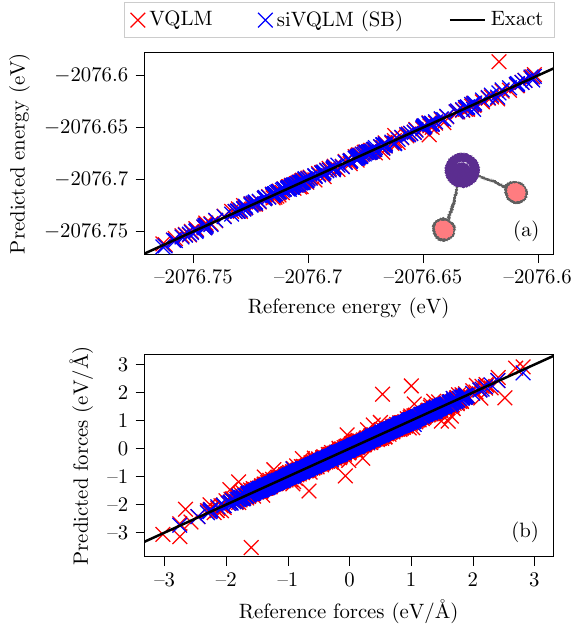}
    \caption{Energy prediction (a) and force prediction (b) for \water.
    The predictions by the generic \ac{VQLM} (red crosses) and the \ac{siVQLM} (blue crosses) including symmetry-breaking layers are compared to the exact energy/forces (black solid line).
    The force prediction in (b) by the generic \ac{VQLM} is given with respect to internal coordinates while the \ac{siVQLM} directly predicts atomic forces with respect to Cartesian coordinates.}
    \label{fig:H2O-results}
\end{figure}

While the \ac{VQLM} and the non-symmetry-breaking \ac{siVQLM} yield prediction accuracy of similar quality, including symmetry-breaking in the \ac{siVQLM} improves both the energy and the force prediction by one order of magnitude.
Both \acp{siVQLM} yield almost the same prediction accuracy for energy and forces on both the training and test data set, meaning that the optimized \acp{siVQLM} can generalize well from the training labels to unknown atomic configurations.
In contrast, the original \ac{VQLM} has larger deviations between training and test losses, indicating worse generalization behavior.
Since including symmetry-breaking yields better prediction accuracy, only the results for the original \ac{VQLM} and the symmetry-breaking \ac{siVQLM} are visualized in \cref{fig:H2O-results}.

\subsection{Triatomic dimer composed of two atom types: water}
\begin{figure}[h]
    \centering
    \includegraphics[width=\columnwidth]{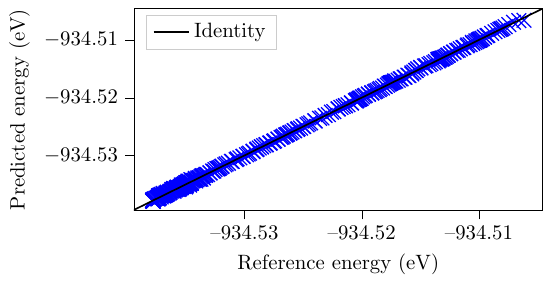}
    \caption{Energy prediction against exact energy labels for \water dimer.
    }
    \label{fig:two_H2O_results}
\end{figure}

\begin{figure}[h]
    \centering
    \includegraphics[width=\columnwidth]{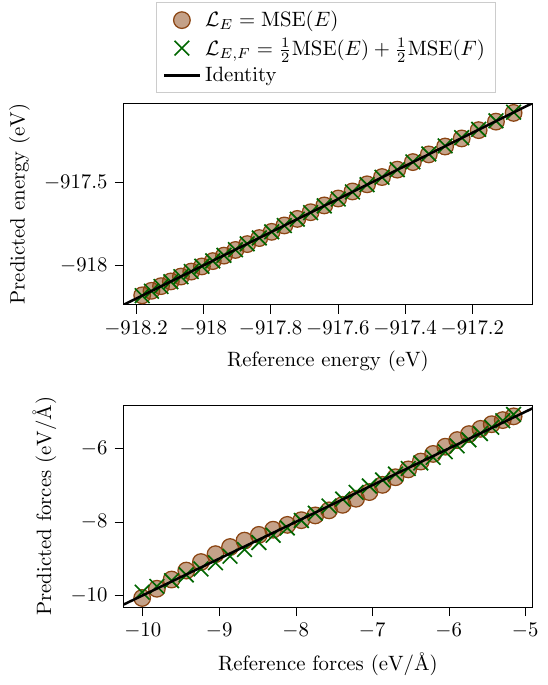}
    \caption{Energy (a) and force (b) prediction for the \water dimer 1D-cut. The effect of excluding (brown circles) and including (green crosses) forces in the training process is compared.
    \label{fig:two_H2O_results_1D}
    }
\end{figure}

Finally, we consider a dimer of triatomic molecules, specifically a water dimer. 
We build the \ac{siVQLM} with $D=30$ encoding and trainable layers, where the latter also includes a symmetry-breaking layer.
Overall, this leads to a model with $N_\text{params}=122$ trainable weights.
The reference energy and force labels are obtained by means of \textit{ab initio} molecular dynamics computation using CPMD~\cite{hutter2005cpmd}.
We train and test the \ac{siVQLM} on $|\mc{A}_\text{train}|=|\mc{A}_\text{test}|=400$ data points respectively for a maximum number of $N_\text{maxiter}=3000$ iterations with the same loss function and optimizer as for the previous examples.
The numerical results are listed in \cref{tab:two_H2O_results}.
\begin{table}[H]
\caption{\label{tab:two_H2O_results}
Numerical results for \water dimer obtained by the \ac{siVQLM} on both the training set $\mc{A}_\text{train}$ and the test set $\mc{A}_\text{test}$.}
\begin{ruledtabular}
\begin{tabular}{lcc}
\multicolumn{1}{c}{\textit{Water dimer}} & $\mc{A}_\text{train}$ & $\mc{A}_\text{test}$ \\
\hline
$\text{MSE}(E)$ (eV$^2$) & $7.26\cdot 10^{-9}$ & $7.45\cdot 10^{-9}$ \\
$\text{MSE}(F)$ (eV$^2$\AA$^{-2}$) & $3.28\cdot 10^{-3}$ & $3.28\cdot 10^{-3}$ \\
\end{tabular}
\end{ruledtabular}
\end{table}
Despite the significant increase in complexity compared to single-molecule cases, the \ac{siVQLM} can successfully predict energy labels for previously unseen chemical configurations.
A visualization of the energy prediction on the test data is given in \cref{fig:two_H2O_results}.
However, the forces predictions still suffer from significant inaccuracies.
This effect is likely due to a lack of expressivity of the model, which limits the overall accuracy with which the potential energy surface can be learned, and -- most importantly -- by the absence of information about force labels in the loss function. Both extensions require heavier computational costs for the classical simulation of the \acp{VQLM} and are left for future investigations. Specifically, the inclusion of forces in the loss function, for instance in the form $\mc{L}=\frac{1}{2}\text{MSE}(E)+\frac{1}{2}\text{MSE}(F)$~\cite{Kiss_2022}, would require the computation of second derivatives from the parametrized quantum circuit if used in combination with a gradient descent training method. 
We nevertheless demonstrate such predicted improvements in a simplified problem instance, by considering a 1D-cut of the potential energy surface for which we only allow for one hydrogen atom to move in one direction while keeping all other coordinates fixed.
First, we train the \ac{siVQLM} on this data subset by only considering the mean-squared-error on the energy prediction as the loss function, i.e., $\mathcal{L}_E=\text{MSE}(E)$.
Next, we additionally include the forces in the loss function by choosing $\mathcal{L}_{E,F}=\frac{1}{2}\text{MSE}(E) + \frac{1}{2}\text{MSE}(F)$, and train the model with the gradient-based ADAM optimizer.
The resulting energy and force predictions are listed in \cref{tab:two_H2O_results_1D} and visualized in \cref{fig:two_H2O_results_1D}. 

\begin{table}[H]
\caption{\label{tab:two_H2O_results_1D}
Numerical prediction results on the test set for the \water dimer 1D-cut for excluding ($\mathcal{L}_{E}$) and including ($\mathcal{L}_{E,F}$) forces in the loss function.}
\begin{ruledtabular}
\begin{tabular}{lcc}
\multicolumn{1}{c}{\textit{Water dimer 1D-cut}} & $\mathcal{L}_{E}$ & $\mathcal{L}_{E,F}$ \\
\hline
$\text{MSE}(E)_\text{test}$ (eV$^2$) & $1.04\cdot 10^{-6}$ & $3.19\cdot 10^{-7}$ \\
$\text{MSE}(F)_\text{test}$ (eV$^2$\AA$^{-2}$) & $3.79\cdot 10^{-3}$ & $4.58\cdot 10^{-4}$ \\
\end{tabular}
\end{ruledtabular}
\end{table}

As it can be seen, by training the \ac{siVQLM} on both energy and force labels we can improve both the energy and force prediction by one order of magnitude.

\section{\label{sec:discussion}Discussion}
In this work, we presented a collection of \acp{VQLM} based on equivariant quantum neural networks respecting relevant sets of molecular symmetries, and applied them to the task of learning force fields.
We started from a four-qubit \ac{siVQLM} for diatomic molecules respecting \sutwo-invariance, which we have later extended to the case of triatomic molecules with two atom types respecting also invariance under permutation of identical atoms.
Moreover, the layout of a \ac{siVQLM} for a dimer system composed of two triatomic molecules with two atom types has also been described.

For the paradigmatic molecules LiH and \water, we found that including underlying symmetries as an inductive bias in \acp{siVQLM} yields an improvement in the prediction accuracy of about one order of magnitude in comparison to generic \acp{VQLM} built without assuming any specific structure in the data. In fact, the symmetry-informed models produce smoother potential energy surfaces, leading to better gradients, and appear to be more resilient to the presence of noise in the training labels.
Moreover, our \acp{siVQLM} can directly predict forces with respect to Cartesian coordinates -- as required in molecular dynamics -- without any conversion to and from internal coordinates or symmetry functions. 
Interestingly, we also observed that, for the more complex learning tasks, an active but controlled breaking of the symmetries in the design of the \acp{siVQLM} leads to improved performances, by allowing more effective training.

In summary, our proposed architecture represents a necessary key step towards a broader application of geometric QML workflows to molecular force fields generation. Further improvements could be achieved by adopting a more systematic fragmentation approach~\cite{behler2007generalized} -- of which our last example for a water dimer represents a minimal component -- with different sub-networks associated with different chemical species or system portions. An even greater impact could be expected from a tighter integration of classical and QML methods, for instance by using symmetry-invariant quantum circuits to produce novel sets of atomic and molecular descriptors to be later analyzed by classical neural networks. Indeed, the latter can easily be operated and trained at scale, while quantum maps may be able to capture classically inaccessible forms of correlation.

\section{Acknowledgments}
The authors thank Johannes Jakob Meyer for fruitful discussions about equivariant models. 
Simulations were performed with computing resources granted by RWTH Aachen
University under projects 5422 and 6337. 
This research was supported by the NCCR MARVEL, a National Centre of Competence in Research, funded by the Swiss National Science Foundation (grant number 205602), and by the NCCR SPIN, a National Centre of Competence in Research, funded by the Swiss National Science Foundation.
I.L. was supported by the German Academic Exchange Service (DAAD) within the \textit{IFI}-scholarship program.
O.K. is supported by CERN through the CERN Quantum Technology Initiative. 
IBM, the IBM logo, and ibm.com are trademarks of International Business Machines Corp., registered in many jurisdictions worldwide. Other product and service names might be trademarks of IBM or other companies. The current list of IBM trademarks is available at \url{https://www.ibm.com/legal/copytrade}.

\begin{acronym}
    \acro{VQLM}{variational quantum learning model}
    \acro{siVQLM}{symmetry-invariant VQLM}
    \acro{H2O}{water}
    \acro{LiH}{lithium hydride}
\end{acronym}

\bibliography{bibliography_paper}
\end{document}